\documentclass{llncs}
\usepackage{makeidx}  

\newcommand{\be}{\begin{equation}}
\newcommand{\bea}{\begin{eqnarray} \nonumber}
\newcommand{\ee}{\end{equation}}
\newcommand{\eea}{\end{eqnarray}}

\begin{document}
\pagestyle{headings}  

\mainmatter              
\title{On the probabilistic approach to the random  satisfiability problem}
\titlerunning{Random K-Sat}  
%
\author{Giorgio Parisi}
\authorrunning{Giorgio Parisi}   
%
%
\pagestyle{empty}
\institute{Dipartimento di Fisica, Sezione INFN, SMC and   UdRm1 of INFM,\\
Universit\`a di Roma ``La Sapienza'', \\
Piazzale Aldo Moro 2, I-00185 Rome (Italy)\\
\email{giorgio.parisi@roma1.infn.it},\\ WWW home page:
\texttt{http:chimera.roma1.infn.it}
}

\maketitle              

\begin{abstract}
In this note I will review some of the recent results that have been obtained in the probabilistic approach to the 
random satisfiability problem.  At the present moment the results are only heuristic. In the case of the random  
3-satisfiability problem a phase transition from the satisfiable to the unsatisfiable phase is found at $\alpha=4.267$.
There are other values of $\alpha$ that separates different regimes and they will be described in details.
In this context the properties of the survey decimation algorithm will also be discussed.
\end{abstract}
\section{Introduction}
Recently many progresses \cite{MPZ,MZ} have been done on the analytic and numerical study of the random K-satisfiability 
problem \cite{COOK,KS,sat,sat0}, using the approach of survey-propagation that generalizes the more old approach based 
on the belief-propagation algorithm \footnote{The belief propagation algorithm (sometimes called ``Min-Sum'') is the the 
zero temperature limit of the ``Sum-Product'' algorithm.  In the statistical mechanics language \cite{MPV} the belief 
propagation equations are the extension of the TAP equations for spin glasses \cite{TAP,primo} and the survey-propagation 
equations are the TAP equations generalized to the broken replica case.} \cite{BP,MPV,factor,MoZ} . Similar results have 
also been obtained for the coloring a of random graph \cite{MPWZ}.

In the random K-sat problem there are $N$ variables $\sigma(i)$ that may be true of false (the index $i$ will sometime 
called a node).  An instance of the problem is given by a set of $M\equiv \alpha N$ clauses.  In this note we will 
consider only the case $K=3$.  In this case each clause $c$ is characterized by set of three nodes 
($i^{c}_{1}$,$i^{c}_{2}$, $i^{c}_{3}$), that belong to the interval $1-N$ and by three Boolean variables 
($b^{c}_{1}$,$b^{c}_{2}$, $b^{c}_{3}$, i.e. the signatures in the clause).  In the random case the $i$ and $b$ variables 
are random with flat probability distribution.  Each clause $c$ is true if the expression
\begin{equation}
E^{c}\equiv (\sigma(i^{c}_{1}) XOR\, b^{c}_{1}) \ OR \ (\sigma(i^{c}_{2}) XOR\, b^{c}_{2}) \ OR \ 
(\sigma(i^{c}_{3}) XOR\, b^{c}_{3})
\end{equation}
is true 
\footnote{When all the $b^{c}$ are false 
$E^{c}=\sigma(i^{c}_{1})\ OR \ \sigma(i^{c}_{2}) \ OR \ \sigma(i^{c}_{3})$ 
while when all the $b^{c}$ are true 
$E^{c}=\overline{\sigma(i^{c}_{1})}\ OR \ \overline{\sigma(i^{c}_{2})} \ OR \ 
\overline{\sigma(i^{c}_{3})}$.
}.

The problem is satisfiable iff we can find a set of the variables $\sigma$ such that all the clauses are true 
(i.e. a legal configuration); in other words we must find a truth value assignment.  
The entropy \cite{MoZ} of a satisfiable problem is the logarithm of the number of the different sets of the $\sigma$ 
variables that make all the clauses true, i.e the number of legal configurations.

The goal of the analytic approach consists in finding for given $\alpha$ and for large values of $N$ the 
probability that a random problem (i.e. a problem with random chosen clauses) is satisfiable.  The $0-1$ 
law \cite{KS,sat0,01} is supposed to be valid: for $\alpha<\alpha_{c}$ all random systems (with 
probability one when $N$ goes to infinity) are satisfiable and their entropy is proportional to $N$ with 
a constant of proportionality that does not depend on the problem.  On the other hand, for 
$\alpha>\alpha_{c}$ no random system (with probability one) is satisfiable.  An heuristic 
argument\cite{MPZ,MZ} suggests that $\alpha_{c}=\alpha ^{*}\approx 4.267$ where $\alpha^{*}$ can be 
computed using the survey-propagation equations defined later.  There is already a proof \cite{FL} that 
the value of $\alpha ^{*}$ computed with the techniques of survey-propagation is a rigorous upper bound 
to $\alpha _{c}$ (the proof has been obtained only for even $K$, the extension to odd $K$ is technically 
difficult).

\section{Results}

Generally speaking we are interested to know not only the number of legal configurations, but also the properties of the set of all legal 
configurations. At this end it is convenient to say that two configurations are adjacent if their Hamming distance is 
less than $\epsilon N$, where $\epsilon$ is a small number.

We can argue that in the limit of large $N$:
\begin{enumerate} 
	 \item In the interval $\alpha<\alpha_{d}\approx 3.86$ the set of all legal configurations is
	 connected, i.e. there is a path of mutually adjacent configurations that joins two configurations
	 of the set.  In this region the belief-propagation equations (to be define later) have an unique
	 solution.

    \item In the interval $\alpha_{d}<\alpha<\alpha_{c}\approx 4.267$ the set of all the legal 
    configurations breaks in an large number of different disconnected regions that are called with 
    many different names in the physical literature \cite{MP1,MP2} (states, valleys, clusters, lumps\ldots).  
    Roughly speaking the set of all the legal configurations can be naturally decomposed into  clusters of 
    proximate configurations, while configurations belonging to different clusters (or regions) are not 
    close. This phenomenon is called in spontaneous replica symmetry breaking in the physical literature. 
    The core of the approach of this note is the analysis of this phenomenon and of the methods used to 
    tame its consequences \footnote{Other models, where this phenomenon is not present, like random 
    bipartite matching can be analyzed in a much simple way, although 15 years have been needed from
    the statements of the main result (i.e. the length of the shortest matching in the infinite $N$ 
    limit is $\zeta(2)$) to the rigorous proof of this fact.}. The precise definition of these regions is 
    rather complex \cite{PARISILH}; roughly speaking we could say that two legal configurations belongs 
    to the same region if they are in some sense adjacent, i.e. they belongs to a different region if 
    their Hamming distance is greater than $\epsilon N$.  In this way the precise definition of these 
    regions depends on $\epsilon$, however it can be argued that there is an interval in $\epsilon$ 
    where the definition is non-trivial and is independent from the value of $\epsilon$: for a rigorous 
    definition of these regions see \cite{TALE,DuMa,CDMM}. The  number of these regions is given by $\exp( 
    \Sigma^{N}(\alpha))$, where $\Sigma^{N}(\alpha)$ is the total complexity; for large $N$ the total 
    complexity is asymptotically given by $\Sigma^{N}(\alpha)= N\Sigma(\alpha)$ where $\Sigma(\alpha)$ is 
    the complexity density. In this interval the belief-propagation equations have 
    many solutions and each of these solution  is associated to a different cluster. The statistical 
     properties of the set of the solutions of the belief-propagation equations can be studied using 
     the belief-propagation equations (to be defined later).

    \item Only in the interval $\alpha_{b}\approx 3.92<\alpha<\alpha_{c}$ there are literals $\sigma$ that are frozen, i.e. they 
    take the same value in all the legal configurations of a region \footnote{The distinction between $\alpha_{d}$ 
    \cite{sat} a 
    and $\alpha_{b}$ \cite{MPZ} is not usually done in the literature and sometimes it is wrongly assumed that 
    $\alpha_{b}=\alpha_{d}$.}. We could say that the frozen variables form the backbone of a 
    region. It is important to realize that a given clause may simultaneously belong to the backbone of one region 
    and not belong to the backbone of an other region.

\end{enumerate}

The arguments run as follow.  Let us start with a given instance of the problem.  We first write the
belief propagation equations.  For each clause that contains the node $i$ (we will use the notation
$c \in i$ although it may be not the most appropriate)  $p_T(i,c)$ is defined to be the probability that the
variable $\sigma(i)$ would be true in absence of the clause $c$ when we average over the set of all 
the legal configuration ($p_F(i,c)=1-p_T(i,c)$ is the probability to be false).  If the node $i^{c}_{1}$ were
contained in only one clause, we would have that 

\bea p_T(i^{c}_{1})= 
u_T(p_T(i^{c}_{2},c),p_T(i^{c}_{3},c),b^{c}_{1},b^{c}_{2},b^{c}_{3}) \equiv u_T(i_{1},c) \ , \\
p_F(i^{c}_{1})=1- u_T(i_{1},c) \ ,
\eea
where $u_T$ is an appropriate function that is defined by the previous relation.
An easy computation shows that when all the $b$ are false, the variable $\sigma(i^{c}_{1})$ must be true if  both  
variable  $\sigma(i^{c}_{2})$ and  $\sigma(i^{c}_{3})$ are false, otherwise it can have any value. Therefore we have in 
this case that
\be
u_T(i,c)=\frac {1}{2-p_F(i^{c}_{2},c) p_F(i^{c}_{3},c)}\ .
\ee
In a similar way, if some of the $b$ variable are true, we should exchange the indices $T$ and $F$ for 
the corresponding variables, i.e., if $b_1^c$ is true, then $u_T(u_{1})$ becomes $u_F(u_{1})$.  Finally 
we have that
\bea
p_T(i,c)={\prod_{d\in i, d\ne c}u_T(i,d) \over Z_{0}(i,c)} \\
p_F(i,c)={\prod_{d\in i, d\ne c}u_F(i,d) \over Z_{0}(i,c)}  \nonumber \\
Z_{0}(i,c)=\prod_{d\in i, d\ne c}u_T(i,d)+\prod_{d\in i, d\ne c}u_F(i,d).
\eea
We note  the previous formulae can be written in a more compact way 
 if we introduce a two dimensional vector $\vec{p}$, with components $p_T$ and $p_F$.
We define the product of these vector 
\be
c_T=a_T \  b_T \ \ \ \ \ c_F=a_F \  b_F , 
\ee
if $\vec{c}=\vec{a} \cdot \vec{b}$. 

If the norm of a vector is defined by
\be
|\vec{a}|=a_T+a_F \ ,
\ee
the belief propagation equations are defined to be
\be
\vec{p}(i,c)={\prod_{d\in i, d\ne c}\vec{u}(i,d) \over |\prod_{d\in i, d\ne c}\vec{u}(i,d)|}\ .
\ee

In total there are $3M$ variables $p_T(i,c)$ and $3M$ equations.  These equations in the limit of large $N$ should have 
an unique solution in the interval $\alpha<\alpha_{d}$ and the solution should give the correct values 
for the probabilities  $p_T(i,c)$. In this interval the entropy (apart corrections that are subleading 
when $N$ goes to infinity) is given by
\be
S=-\sum_{i=1,N} \log(Z_{1}(i)) +2\sum_{c=1,M}\log(Z_{2}(c)) \ .
\ee
Here the first sum runs over the nodes and the second one runs over the clauses; $Z_{2}(c)$ is the 
probability that the clause $c$ would be satisfied in a system where the validity of the clause $c$ is 
not imposed.  One finds that in the case where all the $b$ variables are false
\be
Z_{2}(c)=1-p_F(i^{c}_{1})p_F(i^{c}_{2})p_F(i^{c}_{2}) \ .
\ee
In a similar way $Z_{1}(i)$ is the probability that all we can find a legal configuration containing the site $i$ starting 
from a configuration where the site $i$ is not present and it is given by:
\be
Z_{1}(i)=|\prod_{d\in i}\vec{u}(i,d)|
\ee
The belief propagation equations can also be written as :
\be
{\partial S \over \partial \vec{p}(i,c)}=0 \ .
\ee
The belief-propagation equations can be formally derived by a local analysis by assuming that, in
the set of the legal configurations of the system where all the clauses $c \in i$ are removed, the
variables $\sigma(k)$ that would enter in these clauses are not correlated.  This cannot is not true
for finite $N$, but this statement may be correct in the limit $N \to \infty$ with the appropriate
qualifications.

Generally speaking for a given sample these equations may not have an exact solution, but they do have 
quasi-solutions \cite{P2} (i.e. approximate solutions, where the approximation becomes better and better 
when $N$ goes to infinity \footnote{More precisely if we have $N$ equations $E_{i}[\sigma]=0$ for $i=1,N$ 
a solution $\sigma$ of this system of equation satisfies the condition 
$N^{1}sum_{i=1,N})(E_{i}[\sigma])^{2}=0$; a quasi-solution satisfies the weaker condition 
$N^{1}sum_{i=1,N})(E_{i}[\sigma])^{2}<h(N)$, where $h(N)$ is a function that goes to zero when $N$ goes 
to infinity.  The definition of a quasi-solution depends on the properties of the function $h(N)$ and this 
point must be further investigated: it may turn out at the end that quasi-solutions are not needed.}): 
these equations have been derived using a local analysis that is correct only in the limit $N \to 
\infty$.
 
 In the interval 
 $\alpha_{d}<\alpha<\alpha_{b}$ the variables $p_F$ and $p_T$ are different from 0 and 1; however in
 the region $\alpha_{b}<\alpha<\alpha_{c}$ there solutions (or quasi-solutions) of the belief
 equations have a fraction of the variables $p_F$ and $p_T$ that are equal to 0 or 1.
 
 When the number of solutions of the belief propagation equations is large, 
 the properties of the sets of solutions of the belief propagation equations can be obtained by computing the solution of 
 the survey propagation equations defined as follows.  
 In the general approach in each node we introduce the probability (${\cal P}_{i,c}(p)$) to find a
 solution of the belief-propagation equations with $p_{T}(i,c)=p$.  With some effort we can write down  
 local equations
 for this probability.  These are the full survey equations that allow the computation of the total
 entropy.  
 
This approach is computationally heavy.  As far as the computation of the complexity is concerned, we can 
use a simpler approach, where we keep only a small part of the information contained in the function 
${\cal P}_{i,c}(p)$, i.e. the weight of the two delta function at $p=0$ and $p=1$.  More precisely we 
introduce the quantity $s_T(i,c)$ that is defined as the probability of finding $ p_T(i,c)=1$, in the 
same way $s_F(i,c)$ is the probability of finding $ p_T(i,c)=0$ and $s_{I}(i,c)$ is the probability of 
finding $0< p_T(i,c)<1$.  It is remarkable that it is possible to write closed equations also for these 
probabilities (these equations are usually called the survey propagation equations \cite{MPZ}).

We can use a more compact notation by introducing a three dimensional vector $\vec{s}$ given
by

\be
\vec{s}= \{ s_T,s_I,s_F\} \ .
\ee
Everything works as before with the only difference that we have a three component vector instead of a two component 
vector. 
Generalizing the previous arguments one can introduce the quantity $\vec{u}(i,c)$ that is the value that the 
survey at $i$ would take if only the clause $c$ would be present in $i$ (in other words $\vec{u}(i,c)$ is 
the message that arrives to the site $i$ coming from the clause $c$).
In the case where all the $b$ are false, a simple computation gives 
\be
\vec{u}(i,c) =\{ s_F(i^{c}_{2},c)  s_F(i^{c}_{3},c),\   1- s_F(i^{c}_{2},c)  s_F(i^{c}_{3},c) ,\  0 \} \ . \label{A}
\ee
The formula can generalized as before \footnote{It always happens that the  vector $\vec{u}$ has only one  zero 
component ($u_{T}u_{F}=0$). This fact may be used to further simplify the analysis.} to the case  of different values of $b$. 
One finally finds the survey propagation equations:
\be
\vec{s}(i,c)={\prod_{d\in i, d\ne c}\vec{u}(i,d) \over |\prod_{d\in i, d\ne c}\vec{u}(i,d)|} \ , \label{GUAI}
\ee
where we have defined product in such a way that 
\be
\vec{a}\vec{b}=\{a_T b_T+a_{I}  b_T  +a_T b_{I}, a_{I}  b_{I},  \ a_F\ b_F+a_{I}\  b_F  +a_F\  b_{I} \}
		 . \label{B}
\ee

It is convenient to introduce the reduced complexity ($\Sigma_{R}(\alpha)$), that  counts the number of solutions of the 
belief equations where two different solutions are considered equals if they coincide in the points where the 
beliefs are 0 or 1 \footnote{It is not clear at the present moment if there are different solutions of the belief 
equations that coincide in all the points where the 
probability is 0 or 1: in other words we would like to know if $\Sigma_{R}(\alpha)=\Sigma(\alpha)$, where 
$\Sigma(\alpha)$ counts the total number of solutions of the belief equations.}. In other words two 
solutions of the beliefs equations with an identical backbone enters only once in the  counting that 
leads to the reduced complexity.

If there is an unique solution to the survey propagation equations, it is possible to argue that
the reduced total complexity should be given by
\be
\Sigma_{R}=-\sum_{i=1,N} \ln(Z_{1}(i))+2\sum_{c=1,M}\ln(Z_{2}(c))
\ee
where now the definition of the $Z$'s is changed and it is done using the surveys, not the beliefs:
\be
Z_{1}(i)=\ln(|\prod_{d\in i}\vec{u}(i,d)|), \ \ \ 
Z_{2}(c)=\ln(|\vec{s}(i,c)\vec{u}(i,c)|)
\ee
The reduced complexity $\Sigma_{R}(\alpha)$ it is particularly interesting because it hat been conjecture 
that it should vanishes at 
the critical point $\alpha_{c}$. This allow the computation of the point $\alpha_{c}$. 

 It is interesting that also in this case the survey propagation equations can be written in a simple 
 form:
 \be
 {\partial \Sigma_{R} \over \partial \vec{s}(i,c)}=0 \ .
\ee

 One finally finds that the survey-propagation equations do not have  an unique solution when
 $\alpha>\alpha_{U}\approx 4.36$. The fact is not of  direct importance because  
 $\alpha_{U}>\alpha_{c}$. Indeed in the region $\alpha>\alpha_{c}$ the complexity is negative so that 
 the there are no solutions of the belief-propagation equations associated to the solution of the 
 survey-propagation equation.

It is evident that for $\alpha>\alpha_{c}$ there are no more legal 
configurations and  $\Sigma_{R}(\alpha)$ is not well defined. A negative value $\Sigma_{R}(\alpha)$ can 
be interpreted by saying that the probability of finding a legal configuration goes to zero exponentially 
with $N$. We stress that the entropy density remains finite at $\alpha_{c}$,  the conjecture that $\Sigma_{R}(\alpha)$
vanishes at $\alpha_{c}$ implies that the reduced complexity is captures the number of essentially 
different regions of legal configuration. A priori a finite value $\Sigma_{R}(\alpha_{c})$ cannot be 
excluded.

\section{Methods}

We now show how to obtain the above results on the 
solutions of the belief propagation
equations (and of the survey propagation equations) for a large random system in the limit of large
$N$. These equations are interesting especially in the infinite $N$ limit where the factor graph does not 
contain short loop. For finite $N$ in the random case, (or  in the infinite $N$ limit for a non-random 
case) the belief equations may have solutions, but the properties of these solutions do not represent 
exactly the properties of the systems. If the number of short loops is small, perturbative techniques may be 
used to compute the corrections to the belief equations. If short loops  are very common (e.g. if the 
literals are on the sites of an f.c.c. lattice and the clauses are on faces of the same lattice), it is rather likely 
that the beliefs equations are  useless and they could only used as starting point of much more 
sophisticated approaches.

We attack this problem by studying the solution of the belief propagation equations (and of the
survey propagation equations) on an random infinite tree.  Sometimes the solution is unique, i.e. it
does not depends on the boundary conditions at infinity, and sometimes it is not not unique.  The
statistical properties of the solution of these equations can be studied with probabilistic method. 
One arrives to integral equations that can be solved numerically using the method of population
dynamics \cite{MPZ,MZ,MP1,MP2}.  The numerical solutions of these integral equations can be used to
compute $\alpha_{d}$, $\alpha_{b}$, $\alpha_{c}$, and $\alpha_{U}$.
 
 The generalization of the Aldous construction \cite{ALDOUS,P1} of an infinite rooted tree associated to a 
graph can play the role of a bridge between a finite instance of the problems and the infinite random tree 
where analytic computations \cite{MPZ,MZ,MP1,MP2} are done. For example it could be used to prove the 
existence and the uniqueness of the beliefs and survey propagation equation in the appropriate intervals.

In this way one can argue that the 
properties on an infinite random tree are relevant for the behaviour of a given random system in the 
limit of large $N$.
 
We can check that the results we have obtained in these way for the solution of the belief and
survey propagation equations are correct by computing in an explicit way the solution (when it is
unique) of the equations for a given sample for large $N$  (e.g $N=10^{4}-10^{6}$). For example we 
may compare the distribution of the beliefs or of the surveys in a large system  with the one 
obtained by solving the integral equations for the probabilities: the agreement is usually very good.

In the same spirit the validity of the result for $\alpha_{d}$ may be checked by studying the
convergence of the iterative procedure for finding the solution of the belief-propagation equations
on a given large problem One finds that just at $\alpha_{d}$ the iterative procedure for finding a
solution does not converge anymore and this is likely a sign of the existence of many solutions to
the belief-propagation equations.  In a similar way we can check the correctness of $\alpha_{U}$.

\section{Survey decimation algorithm}
The survey decimation algorithm has been proposed \cite{MPZ,MZ,BMWZ,BMZ,P3} for finding the solution of 
the random K-satisfiability problem \cite{KS,sat,sat0}.

 We start by solving the survey propagation equation. If a survey ($\vec{s}(i)$) is very near to $(1,0,0)$ (or to 
 $(0,0,1)$) in most of the legal solutions of the beliefs equations (and consequently in the legal configurations) the 
 corresponding local variables will be true (or false).

The main step in the decimation procedure consists is starting from a problem with $N$ variables and to consider a 
problem with $N-1$ variables where $\vec{s}(i)$ is fixed to be true (or false).  We denote
\be
\Delta(i)=\Sigma^{N}-\Sigma^{N-1} \ .
\ee
If $\Delta(i)$ is small, the second problem it is easier to solve: it has nearly the same number of solutions 
of the belief equations 
and one variable less. (We assume that the complexity can be computed by solving the survey propagation
equations).

The decimation algorithm proceeds as follows.  We reduces by one the number of variables choosing the node $i$ in the 
appropriate way, e.g. by choosing the node with minimal $\Delta(i)$.  We recompute the solutions of the survey 
equations and we reduce again the number of variables.  At the end of the day two things may happen: 
\begin{enumerate}
\item We arrive to a 
negative complexity (in this case the reduced problem should have no solutions and we are lost),
\item The denominator in equation  (\ref{GUAI}) becomes zero, signaling the presence of a contradiction 
(and also in this case we are lost),
\item The non-trivial solution of the survey equation disappears.  If this 
happens the reduced problem is now easy to be solved. 
\end{enumerate}

The quantity $\Delta(i)$ may be estimated analytically so that it is possible to choose the variable with 
minimal $\Delta(i)$.  A careful analysis of the results for large, but finite $N$ \cite{P3} shows that 
the algorithm works in the limit of infinite $N$ up to $\alpha_{A}\approx 4.252$, that is definite less, 
but very near to $\alpha_{c}$.

Unfortunately at the present moment this result for $\alpha_{A}$ can be obtained only 
analyzing how the argument works on a finite sample and we are unable to write down integral 
equations for the probability distributions of the solution of the survey propagation equations. 
This drawback leads to the impossibility of computing analytically $\alpha_{A}$: it is a rather 
difficult task to understand in details why for $\alpha_{A}$ it so near to $\alpha_{c}$. It is 
interesting to note that for $\alpha<\alpha_{A}$ the survey decimation algorithm takes a time that 
is polynomial in $N$ and using a smart implementation the time is nearly linear in $N$.
It is important to stress that  survey algorithm is an incomplete search procedure which
may not be able to find any solution to a satisfiable instance.
This actually happens with a non-negligible
probability e.g. for sizes of the order of a few thousands of
variables also when $\alpha<\alpha_{A}$, however in the limit $N \to \infty$, it should work as soon 
$\alpha<\alpha_{A}$.

In the interval $\alpha_{A}<\alpha<\alpha_{c}$ the decimation procedure leads to a regime of negative 
complexity so that the algorithm does not work.  Unfortunately there is no analytic computation of 
$\alpha_{A}$.  It is likely that the fact that $\alpha_{U}=4.36$ is not far from $\alpha_{c}$ is related 
to the fact the $\alpha_{c}-\alpha_{A}$ iss small.

It would be very interesting to 
understand better if such a relation is true and to put  in a quantitative form. In this regard it would 
be be important to study the $K$ dependence of $\alpha_{c}-\alpha_{A}$ and 
$\alpha_{U}-\alpha_{c}$. This analysis may give some hints why  $\alpha_{c}-\alpha_{A}$  is so small 
in 3-SAT.

\section{Conclusions}
Similar problems have been studied by physicists in the case of infinite range spin glasses \cite{MPV}: here the 
problem consists in finding the minimum  $E_{J}$ of the quantity:
\be
H_{J}[\tau]\equiv \sum_{i,k=1,N}J_{i,k}\tau_{i}\tau_{k}
\ee
where the minimum is done respect to the variables $\tau_{i}=\pm  1$ and the $J$ are independent Gaussian 
variables with zero average and variance $N^{-1/2}$. Physical intuition tells us that in the limit $N$ 
goes to infinity  the intensive quantity  
\be
e_{J}=\frac{E_{J}}{N}
\ee
should be (with probability 1) independent from $N$  and it will be denoted by 
$e_{\infty}$.  In 1979 it was argued using the so called replica method (that will be not discussed in 
this note) that $e_{\infty}$ was equal to the maximum of a certain functional $F[q]$, where $q(x)$ is a 
function defined in the interval $[0-1]$.  Later on, 1985 the same results were rederived using 
heuristical probabilistic consideration, similar to those  presented here (but much more complex).  In 
this note we have introduced a hierarchical construction, where three levels (configurations, beliefs, 
surveys) are presents: in the case of spin glasses an infinite number of levels is needed (in the spin 
glass case the survey 
equations do not have an unique solution and we have to consider higher and higher levels of 
abstraction).  Only very recently Talagrand \cite{TALE2}, heavily using Guerra's ideas and results, was able to 
prove that the value for $e_{\infty}$, computed 24 year before, was correct.

The possibility of using these techniques for deriving eventually exact results on the K-SAT problem is a 
very recent one: only a few year ago \cite{MP1,MP2} the previous techniques has been extended to more complex 
spin glass models where the matrix $J$ is sparse (it has an average number of elements per raw that does 
not increase with $N$).  

The field is very young and rigorous proofs of many steps of the construction (also those that would 
likely be relatively simple) are lacking.  We only know, as a consequence of  general theorems, that 
this methods give an upper bound to the value of $\alpha_{c}$, and this upper bound should be computed by 
maximizing an appropriate functional of the probability distribution of the surveys.  This upper bound is 
rigorous one \cite{FL}: it essentially use positivity arguments (the average of a non-negative function is 
non-negative) in a very smart way and it does not depend on the existence or uniqueness of the solutions 
of the equations for the probability distribution of the survey.  On the contrary the way we have 
followed to compute this upper bound (i.e. $\alpha^{*}$) require some extra work before becoming fully 
rigorous.  I stress that this upper bounds and the Talagrand's exact result do not need in any way  
considerations on the solutions of the survey propagation equations (or of their generalization) on a 
finite sample.  The survey propagation equations are crucial for giving an intuitive image of the 
situation (i.e. at a metaphoric level) and for constructing the survey decimation algorithm.  The 
heuristic derivation could have been done using the replica method, where survey propagation equations 
are never mentioned, but the argument is much more difficult to follow and to transform in a 
rigorous one.

The other results come from empirical (sometimes very strong) numerical evidence and from heuristic 
arguments.  For example at my knowledge there is no proof that the integral equations for the probability 
of the surveys (or of the beliefs) have an unique solution and that the population dynamics algorithm 
converges (to that unique solution).  Proofs in this direction would be very useful and are a necessary 
step to arrive to a rigorous quantitative upper bounds and eventually exact results.  On the other hand 
the proof of existence of an unique solutions (or quasi-solutions) of the surveys (or beliefs) 
propagation equations in the large $N$ limit is lacking  for any value of $\alpha$, although the 
analysis with the population dynamics (whose precise mathematical properties have to be clarified) tell 
us which should the maximum values ($\alpha_{U}$ and $\alpha_{b}$ respectively, below which these 
uniqueness properties hold. Many steps have to be done, but a rigorous determination of $\alpha_{c}$ 
seems to be a feasible task in a not too far future.


\begin{thebibliography}{55}
%
\bibitem {MPZ} M. M{\'e}zard, G. Parisi and R. Zecchina,  Science \textbf{297}, 812 (2002).
\bibitem {MZ} M. M{\'e}zard  and R. Zecchina  {\sl The random K-satisfiability problem: from an analytic solution
to an efficient algorithm} cond-mat 0207194.

 \bibitem{COOK} S.A. Cook, D.G. Mitchell, {\sl Finding Hard Instances
of the Satisfiability Problem: A Survey}, In: \emph{Satisfiability Problem:
Theory and Applications}. Du, Gu and Pardalos (Eds).  DIMACS Series in
\emph{Discrete Mathematics and Theoretical Computer Science}, Volume 35,
(1997)

\bibitem{KS} S. Kirkpatrick, B. Selman, \emph{Critical Behaviour in the
satisfiability of random Boolean expressions}, Science 264, 1297 (1994)

\bibitem{sat} Biroli, G., Monasson, R.  and Weigt, M. \emph{A Variational description of 
the ground state
structure in random satisfiability problems,} {\it Euro. Phys. J. }
{\bf B 14}  551 (2000),

\bibitem {sat0}Dubois O.  Monasson R., Selman B. and   Zecchina R. 
(Eds.), \emph{Phase Transitions in Combinatorial Problems},
Theoret. Comp. Sci. 265, (2001), G. Biroli, S. Cocco, R. Monasson, Physica A 306, 381 (2002). 

 \bibitem{MPV} M\'ezard, M., Parisi, G.  and Virasoro, M.A.\emph{ Spin Glass Theory and Beyond}, World 
Scientific, Singapore, (1987).

\bibitem{TAP} D.J. Thouless, P.A. Anderson and R. G. Palmer, \emph{Solution of a `solvable' model,} Phil.  Mag.  35, 593
(1977)

\bibitem{BP} J.S. Yedidia, W.T. Freeman and Y. Weiss,
\emph{Generalized Belief Propagation}, in {\it Advances in Neural Information
Processing Systems 13} eds. T.K. Leen, T.G. Dietterich, and V. Tresp,
MIT Press 2001, pp. 689-695.

\bibitem{factor} F.R. Kschischang, B.J. Frey, H.-A. Loeliger, \emph{Factor Graphs and the Sum-Product Algorithm}, {\it IEEE
Trans.  Infor.  Theory} {\bf 47}, 498 (2002).

\bibitem{MoZ} Monasson, R. and Zecchina, R. \emph{Entropy of the K-satisfiability problem}, {\it Phys.  Rev.  Lett.} {\bf 76}
3881--3885 (1996).

\bibitem{MPWZ}  R. Mulet, A. Pagnani, M. Weigt, R. Zecchina {\it Phys. Rev. Lett.} {\bf 89}, 268701 (2002).

\bibitem{primo} C. De Dominicis and Y. Y. Goldschmidt: {\it Replica symmetry breaking in 
finite connectivity systems: a large connectivity expansion at finite and zero 
temperature}, J. Phys. A (Math. Gen.)  {\bf 22}, L775 (1989).
 
\bibitem{MP1}  M. M\'ezard and G. Parisi:  Eur.Phys. J. B {\bf 20} (2001) 217;

\bibitem{MP2} M. M\'ezard and G. Parisi: {\it `The cavity method at zero temperature'}, cond-mat/0207121 (2002) to 
appear in J. Stat.  Phys. 

\bibitem{01}  O. Dubois, Y. Boufkhad, J. Mandler, \emph{Typical random
3-SAT formulae and the satisfiability threshold}, in {\it Proc. 11th
ACM-SIAM Symp. on Discrete Algorithms}, 124 (San Francisco, CA, 2000).

\bibitem{FL} S. Franz and M. Leone, \emph{Replica bounds for optimization problems and diluted spin systems},
cond-mat/0208280.
\bibitem{PARISILH}  G. Parisi, \emph{Glasses, replicas and all that} cond-mat/0301157 (2003).
\bibitem{CDMM} S. Cocco, O. Dubois, J. Mandler, R. Monasson. 
     Phys. Rev. Lett. {\bf 90}, 047205 (2003).
     
\bibitem{TALE} M. Talagrand, \emph{Rigorous low temperature results for the
p-spin mean field spin glass model}, {\it Prob. Theory and Related
Fields} {\bf 117}, 303--360 (2000).

\bibitem{DuMa} Dubois and Mandler, FOCS 2002, 769.

\bibitem{ALDOUS} D. Aldous, \emph{The zeta(2) Limit in the Random Assignment
Problem,} Random Structures and Algorithms {\bf 18} (2001) 381-418.

\bibitem{P1}
   G. Parisi: cs.CC/0212047 \emph{On local equilibrium equations for clustering states} (2002).
 \bibitem{P2} G. Parisi: cs.CC/0212009 \emph{On the survey-propagation equations for the random K-satisfiability 
problem}  (2002).
  
\bibitem{BMWZ} A. Braustein, M. Mezard, M. Weigt, R. Zecchina:
cond-mat/0212451
   \emph{ Constraint Satisfaction by Survey Propagation} (2002).

\bibitem{BMZ} A. Braunstein, M. Mezard, R. Zecchina; cs.CC/0212002
\emph{Survey propagation: an algorithm for satisfiability} (2002).


\bibitem{P3} G. Parisi: cs.CC/0301015 \emph{Some remarks on the survey decimation algorithm for K-satisfiability} (2003). 
\bibitem{TALE2}M. Talagrand, private communication (2003).
\end{thebibliography}
\end{document}